\documentclass[twocolumn, secnumarabic,  floatfix, nofootinbib, nobalancelastpage, longbibliography]{revtex4-1}

\def\preprintlinklocation{http://dft.uci.edu + PAPER REF}

\usepackage{amsmath}
\usepackage{physics}
\usepackage{amssymb}
\usepackage{float}
\usepackage[dvipsnames]{xcolor}

\definecolor{TITLECOL}{rgb}{0.05,0.25,0.85}
\definecolor{CONTENTSCOL}{rgb}{0.1,0.2,0.7}
\definecolor{URLCOL}{rgb}{0,0.52,0.83}
\definecolor{LINKCOL}{rgb}{0.05,0.5,0}
\definecolor{CITECOL}{rgb}{0.25,0,0.48}
\definecolor{SECOL}{rgb}{0.07,0.31,0.80}
\definecolor{SSECOL}{rgb}{0.26,0.19,0.75}

\newcommand{\coloredtitle}[1]{\title{\textcolor{TITLECOL}{#1}}}
 
\usepackage{graphicx}

\def\PublicationsLink{http://dft.uci.edu/publications.php}
\def\preprintlink{ \href{\preprintlinklocation}{\TitleOfPaper} }
\def\preprinttext{~}

\usepackage{fancyhdr}
\makeatletter
\fancypagestyle{titlepage}
{	\lhead{\textsc{\preprinttext\ ~ \href{\PublicationsLink}{~}}}
	\chead{}
	\rhead{\preprintlink}
	\lfoot{}}
\makeatother
\def\preprintlink{ 
	\href{\preprintlinklocation}
        {
~}
	}

\definecolor{Green}{rgb}{0.016,0.627,0}
\definecolor{Plum}{rgb}{0.17,0,0.45}
\definecolor{LBlue}{rgb}{0,0.34,0.45}
\definecolor{Sepia}{rgb}{0.37,0.17,0.02}
\definecolor{BurntOrange}{rgb}{0.78,0.39,0}

\usepackage{hyperref}
\hypersetup{
    breaklinks=true,
    bookmarksopen=true,
    bookmarksnumbered=true,
    colorlinks=true,
    linkcolor=LINKCOL,
    linktocpage=true,
    citecolor=CITECOL,
    filecolor=magenta,      
    urlcolor=URLCOL,
}

\def\bea{\begin{eqnarray}}
\def\eea{\end{eqnarray}}
\def\ben{\begin{equation}}
\def\een{\end{equation}}
\def\benu{\begin{enumerate}}
\def\enu{\end{enumerate}}

\def\bei{\begin{itemize}}
\def\eei{\end{itemize}}
\def\beit{\begin{itemize}}
\def\eit{\end{itemize}}
\def\benu{\begin{enumerate}}
\def\enu{\end{enumerate}}

\def\n{n}

\def\sss{\scriptscriptstyle\rm}

\def\1var{(\bx_1...\bx\N)}

\def\br{{\bf r}}

\def\bx{{x}}

\def\bj{{\bf j}}

\def\x{_{\sss X}}

\def\N{_{\sss N}}

\def\Bohr{^{\rm Bohr}}

\def\LDA{^{\rm LDA}}

\def\GEA{^{\rm GEA}}

\def\GGA{^{\rm GGA}}
\def\GE{^{\rm GE}}

\def\B88{^{\rm B88}}

\def\TF{^{\rm TF}}

\def\sph_int{ {\int d^3 r}}

\usepackage{epstopdf}

\usepackage{graphicx}
\usepackage{amsmath}
\usepackage{threeparttable}
\usepackage[normalem]{ulem}
\usepackage{array,multirow}
\usepackage{appendix}
\usepackage{epstopdf}
\usepackage{booktabs}
\usepackage{natbib}
\usepackage{feynmp}
\usepackage{threeparttable}

\graphicspath{{f/}}

\definecolor{SPECOL}{rgb}{0,0.47,0.01}
\definecolor{QUOCOL}{rgb}{0,0,0.2}
\definecolor{SHDCOLb}{rgb}{0.69,0.4,0.1}

\definecolor{SPEQ}{rgb}{0.01,0.4,0.05} %

\definecolor{SPEQv}{rgb}{0.45,0.05,0.45} %

\definecolor{SPEQb}{rgb}{0.01,0.1,0.65} %

\definecolor{SPEQr}{rgb}{0.57,0.05,0.1} %

\def\bay{\begin{array}}
\def\eay{\end{array}}
\def\bit{\begin{itemize}}
\def\beit{\begin{itemize}}
\def\eit{\end{itemize}}

\def\ln{\text{ln} }

\def\floor{\text{floor} }

\def\dd{~ \rotatebox{320}{\hspace{-5pt}\vbox to 5 pt {\hspace{-5pt} \hbox to 5pt {$\cdots$}}}\!\! }

\def\Eqref#1{Eq.~\eqref{#1}}

\definecolor{darkgreen}{rgb}{0.0,0.66,0.0} 
\definecolor{darkred}{rgb}{0.80,0.0,0.0} 
\definecolor{darkblue}{rgb}{0.00,0.0,0.6}

\newcommand\allHL[2]{\textcolor{black}{\sout{}{#2}}}

\begin{document}


\sf 
\coloredtitle{Leading correction to the local density approximation for exchange in large-$Z$ atoms}
\author{\color{CITECOL} Nathan Argaman}
\affiliation{Department of Physics, Nuclear Research Center---Negev, P.O. Box 9001, Be'er Sheva 84190, Israel; argaman@mailaps.org}
\author{\color{CITECOL} Jeremy Redd}
\affiliation{Department of Physics, Utah Valley University, Orem, UT 84058, USA}
\author{\color{CITECOL} Antonio C. Cancio}
\affiliation{Department of Physics and Astronomy, Ball State University,
Muncie, IN 47306, USA}
\author{\color{CITECOL} Kieron Burke}
\affiliation{Departments of Physics and Astronomy and of Chemistry, 
University of California, Irvine, CA 92697,  USA}
\date{4 October, 2022}
\begin{abstract}
The large-$Z$ asymptotic expansion of atomic energies has been
useful in determining exact conditions for corrections to the
local density approximation in density functional theory.
The correction for exchange is fit well with a leading $Z \ln Z$
term, and we find its coefficient numerically. 
The gradient expansion approximation also has such a
term, but with a smaller coefficient.
Analytic results in the limit of vanishing interaction with hydrogenic orbitals (a Bohr atom) lead to the 
conjecture that the coefficients are 
precisely 2.7 times larger than their gradient expansion counterparts, 
yielding an analytic expression for the exchange-energy correction which is accurate to $\sim 5\%$ for all $Z$.
\end{abstract}


\maketitle
\def\floor#1{{\lfloor}#1{\rfloor}}
\def\sm#1{{\langle}#1{\rangle}}
\def\dis{_{disc}}
\newcommand{\Z}{\mathbb{Z}}
\newcommand{\R}{\mathbb{R}}
\def\w{^{(0)}}
\def\w{^{\rm WKB}}
\def\II{^{\rm II}}
\def\hd#1{\noindent{\bf\textcolor{red} {#1:}}}
\def\hb#1{\noindent{\bf\textcolor{blue} {#1:}}}
\def\eps{\epsilon}
\def\ew{\epsilon\w}
\def\ej{\epsilon_j}
\def\upet{^{(\eta)}}
\def\ejeta{\ej\upet}
\def\tjeta{\tj\upet}
\def\bej{{\bar \epsilon}_j}
\def\ewj{\epsilon\w_j}
\def\tj{t_j}
\def\vj{v_j}
\def\F{_{\sss F}}
\def\xt{x_{\sss T}}
\def\sc{^{\rm sc}}
\def\al{\alpha}
\def\ae{\al_e}
\def\bj{\bar j}
\def\bz{\bar\zeta}
\def\eq#1{Eq.\, (\ref{#1})}
\def\cN{{\cal N}}

For almost a century, the non-relativistic semiclassical expansion 
of the total binding energy of atoms \cite{E88a} has guided the development
of density functional approximations, beginning with 
Thomas-Fermi (TF) theory \cite{T27,F28a} and the 
local density approximation (LDA) for exchange \cite{B29a,D30}.
In the seventies, Lieb and Simon proved \cite{LS77} that the dominant term
in that expansion is given exactly by TF theory, 
and in the eighties Schwinger and Englert showed explicitly that the LDA
recovers the dominant term for the atomic exchange 
energy \cite{S81a,ES82a,ES85}.  Recent analytic and
numerical evidence shows the same is
true for atomic correlation energies \cite{BCGP16,CCKB18}.

For exchange, recent focus has been on the
leading correction to LDA \cite{PCSB06,EB09a}, see Fig.~1.  Most modern
generalized gradient approximations (GGAs) --- the starting point of most modern exchange-correlation approximations --- yield a well-defined correction that can be compared to atomic data for large $Z$.  The popular approximations known as PBE \cite{PBE96} and B88 \cite{B88} both yield highly accurate approximations to this term for atoms, which are about double that of the gradient-expansion approximation \cite{Kc57a,DG90} (GEA), yielding some of the insight behind PBEsol \cite{PRCV08}.  The behavior for large $Z$ has been built into several recent non-empirical approximations (SCAN \cite{SRP15}, APBE \cite{CFLD11}, acGGA \cite{CCKB18}).

The original works \cite{PCSB06,EB09a} on expanding the beyond-LDA exchange energy for atoms, 
\ben
\Delta E\x=E\x-E\x\LDA \, , \label{eq:beyond_LDA}
\een
used simple powers of $Z^{1/3}$, based on the scaling behavior of the gradient expansion \allHL{}{for the
slowly-varying electron gas.}
Here we provide three lines of evidence for the existence of a $Z\ln Z$ contribution
{, showing that the} 
analytic forms used as `exact conditions' are likely incorrect, and should be replaced by those suggested below.  
\allHL{}{Thus, the current work not only contributes to the very long-standing search for the
expansion of the energy of atoms in mathematical physics, but also provides
a crucial correction to exact conditions which are built into
the latest modern density functional approximations, used throughout condensed matter physics, materials science, and chemistry.}

\begin{figure}
\includegraphics[trim=0cm 0cm 0cm 0cm,clip=true,width=1.0\linewidth]
{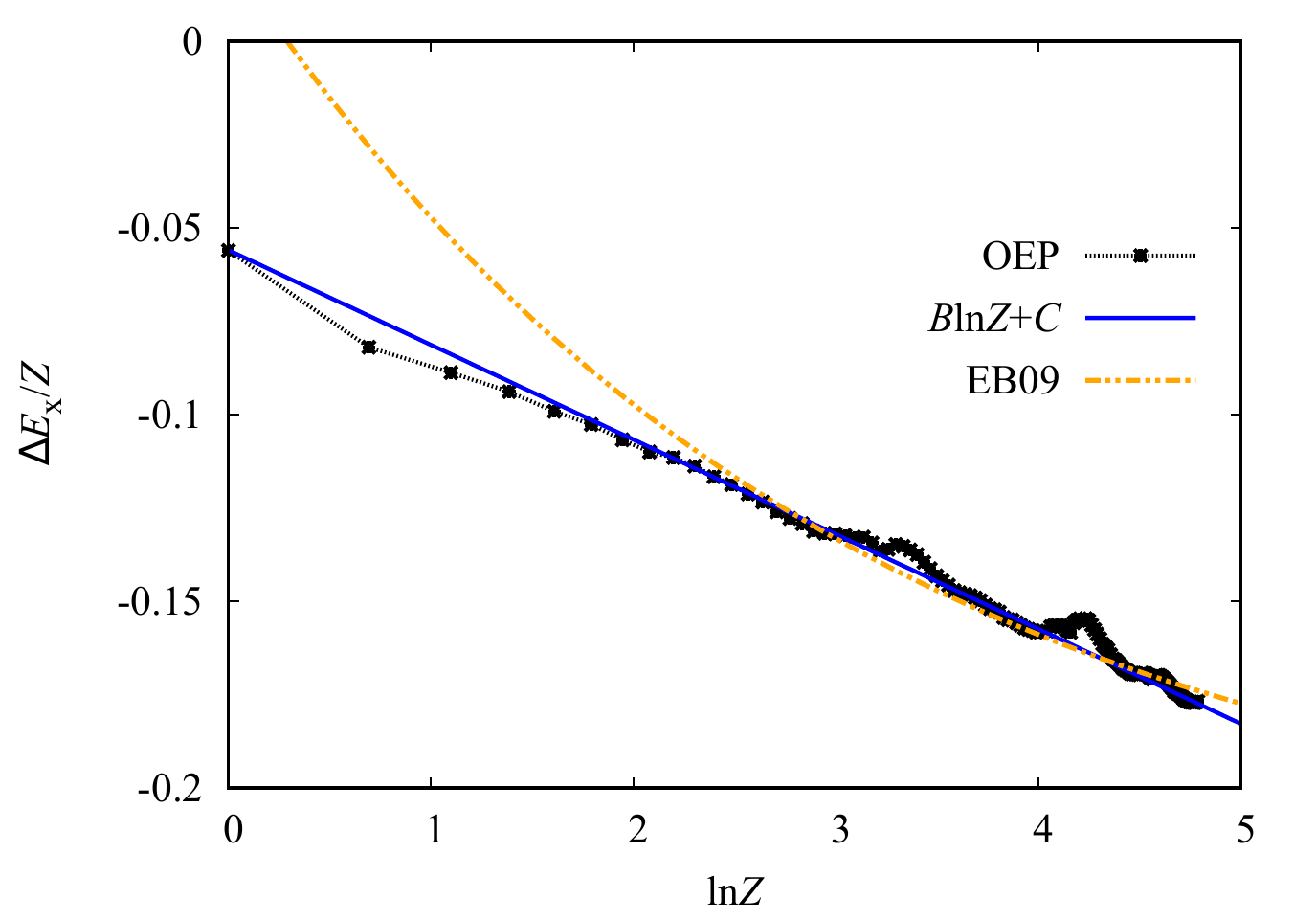}
\caption{Beyond-LDA exchange energy per electron ($\Delta E\x / Z$) of neutral atoms.
The solid blue line is the new $B\ln{Z}+C$ fit described in the text, whereas the orange dashed curve is the fit of Ref.~\cite{EB09a}.
(Hartree atomic units used throughout.)
}
\label{fig:fitlogz}
\end{figure}


Our first line of inquiry consists of evaluating $\Delta E\x /Z$ for neutral atoms up to $Z=120$, using the optimized effective potential (OEP).  
These data are plotted 
versus $\ln Z$ in
Fig.~1, and a straight line gives 
a significantly better fit than Ref.~\cite{EB09a}. 

A second direction shows analytically that 
applying the GEA to the TF density profile for an atom \cite{LCPB09}  
produces a $\ln Z$ divergence near the nucleus, but its
coefficient is less than half the slope of the fit in Fig.~1, reflecting the aforementioned discrepancy with GGAs.

A third direction is a study of the Bohr atom \cite{HL95}, in which the electron repulsion is infinitesimal and the orbitals are hydrogenic.  Exchange energies were calculated analytically for such atoms with up to 22 closed shells \cite{SOLR11a}. 
Fitting these, as well as the LDA exchange energies, 
gives a $Z\ln Z$ coefficient larger than that of neutral, interacting-electron atoms.
In GEA, the cusp
where the Bohr-atom TF density abruptly vanishes also contributes.  
Overall, the $Z\ln Z$ coefficient is here 2.7 times larger than in GEA.
Assuming that ratio is true for all atoms 
explains the data of Fig.~1.



Our first step is a detailed analysis of Fig.~1.
Three candidates for the leading 
correction to LDA are: a term proportional to $Z$~\cite{EB09a}, the $Z\ln Z$ dependence suggested by the GEA,
and a term proportional to $Z^{4/3}$, which appears in the oscillations across the periodic table \cite{ESc85}.
The general form
\ben
     \Delta E\x / Z \approx -A'\ Z^{1/3} -B \ln Z - C - D Z^{-1/3}
\label{eq:postldafit}
\een 
enables a discussion of all these possibilities.

We use the OPMKS code~\cite{ED99} to calculate $E\x$ with the OEP and $E\x\LDA$ \allHL{}{with the spin-dependent LDA of \cite{PW92}}, for non-relativistic neutral atoms up to $Z=120$, extending an earlier data set \cite{BCGP16}.
%
To avoid bias, we ignore the large numbers of highly correlated
data points across subshells, 
keeping only atoms with closed subshells\allHL{}{, grouped as follows: He and}
the alkaline earths (s), \allHL{}{the remaining} noble gases (p), group 12 metals (d), and closed f-shell atoms.
There are 20 such atoms for $Z \leq 120$, but 
we 
exclude the first element of each group ($Z=2$, 10, 30, and 70), as
these are most strongly affected by oscillations in $Z$ \cite{BCGP16}.

To generate a set of competing models for our data, we vary a 
subset of  coefficients in Eq.~(\ref{eq:postldafit}), holding
the others to zero, and find the coefficients and their standard errors 
from nonlinear regression using the Levenberg-Marquardt method \cite{PTVF07}.
These are shown in Table~\ref{table:fits}, 
listed in order of the number of parameters,
with data entries for zeroed out coefficients left blank.  
The final column shows the reduced $\chi^2$ of the
fit, i.e., the sum of the squared errors per degree of freedom,
$\chi^2_{\rm red} = \sum_{i=1}^n 
\left( \frac{\delta_i}{\sigma} \right)^2 / (n-m)$.  
Here $\delta_i$ is the difference between the two sides of 
\Eqref{eq:postldafit} for the $i$th value of $Z$, the standard error 
$\sigma$ has been set to 1 mHa for simplicity, and $m$ is the number 
of free parameters in the fit.

\begin{table}
\begin{ruledtabular}
\renewcommand{\arraystretch}{1.1}
\begin{tabular}{|r|l|l|l|l|l|}
 & $A'$ & $B$ & $C$ & $D$ & $\chi^2_{\rm red}$ \\ 
\hline
1  &             &              & 0.153(6)   &            & 560  \\
2  &             &              & 0.2138(34) & -0.205(11) & 22.1 \\
3  &             & 0.02464(26)  & 0.0590(10) &            & 0.91 \\
4  &             & 0.0256(14)   & 0.053(9)   & ~0.008(12) & 0.95 \\
5  & ~0.0007(15) & 0.0239(16)   & 0.0592(11) &            & 0.96 \\
6  & ~0.0128(9)  &              & 0.134(5)   & -0.098(7)  & 1.3  \\
7  & -0.007(8)   & 0.039(16)    & 0.01(5)    & ~0.06(7)   & 0.98 \\
\end{tabular}
\end{ruledtabular}
\caption{Coefficients of various fits of $\Delta E\x/Z$ in 
         Eq.~(\ref{eq:postldafit}), with ``missing" coefficients fixed
         at zero.  $\chi^2_{\rm red}$ {quantifies} the errors of the fit {as described in the text}.
        Standard errors in the coefficients are given in parenthesis.}
\label{table:fits} 
\end{table}

For the first (and worst) two forms, $\Delta E\x \propto Z$ 
is the leading order,
as in Ref.~\cite{EB09a}.
The logarithmic fit, line 3, has the smallest errors in
coefficients and the best $\chi_{\rm red}$. 
This fit does remarkably well also \textit{outside} the range of $Z$ fitted, even down to hydrogen, as seen in Fig.~\ref{fig:fitlogz}.

The remaining fits have additional free parameters.
A $Z^{1/3}$ term (fits 5 and 7) slightly degrades \allHL{}{ the quality of} the fit,
\allHL{}{in the sense that $\chi_{\rm red}$ increases ($n-m$ decreases more than $\sum_i \delta_i^2$), and the standard error of the $A'$ coefficient is larger than its absolute value, suggesting it should be set to zero \cite{EB09a}.} 
A term proportional to $Z^{-1/3}$ \allHL{also worsens the 
fit}{is likewise ineffective} (fits 4 and 7).
Fit 6, using only powers of $Z^{1/3}$ without a logarithmic term,
\allHL{also}{}%
results in a somewhat larger $\sum_i \delta_i^2$ despite the larger number of free parameters.

An asymptotic series should increase in accuracy as $Z$ increases, so
we refit models to a more restricted set of data: first by dropping a second element of each group
(12 atoms), and then a third (9 atoms).
For the $\ln{Z}$-leading model, the three fits yield essentially
the same results 
($B=0.0254, 0.0253$ and $C=0.0560, 0.0562$).
For the $Z^{1/3}$ model (fit 6), the coefficients drift noticeably as the data is 
restricted to a smaller range, 
\allHL{this drift is a sign that the behavior of}{and} 
\allHL{}{the fit is poor outside
the range fitted, similar to the EB09 curve in Fig.~1.}
\allHL{}{As a final test, using all data from $Z = 1$ to $120$ indiscriminately yields coefficients for $B$ and $C$ that are statistically indistinguishable from those of fit 3, but with a much higher $\chi^2_{\rm red}$.  The data and details of the fits are given in \cite{SM}.}

Overall, the fits with the $\ln Z$ term as leading order are clearly the most predictive,
\allHL{}{and for best judgement of the asymptotic behavior we choose the 12-atom fit of the $\ln{Z}$ model, which is appropriately weighted to large $Z$ (the 9-atom fit gives larger standard errors for $B$ and $C$ [28]):}
\ben
     \Delta E\x \approx -0.0254 Z \ln{Z} - 0.0560 Z,
     \label{eq:fit3}
\een
which is the curve shown in Fig.~1.
Remarkably, given that $E\x\LDA$ is $-0.2564$ for hydrogen,
this yields $-0.3124$, almost exactly matching the analytic result, $-5/16$.
\allHL{}{That \allHL{such success}{the success of this fit} should in fact be expected of the semiclassical approximation is evident in Figs.~1
and 9 of \cite{BCGP16} and in \cite{BB20}.}
\allHL{}{Before such an asymptotic expansion diverges, the inclusion of the next term will often improve
accuracy by two orders of magnitude \cite{B20,BB20}.}
\allHL{providing}{\Eqref{eq:fit3} thus provides} another example of ``the principle of unreasonable utility of asymptotic estimates'' \cite{S80a}.


Next, we estimate $\Delta E\x$ theoretically.  The LDA exchange energy is given by
\ben
E\x\LDA = - a\x \int d^3r\ \n^{4/3}(\br) \; ,
\label{EX_LDA}
\een
where {$a\x =  3 \left( 3 / \pi \right)^{1/3} \! /4$} \cite{B29a,D30}, and insertion of the TF density \cite{LCPB09} into this expression directly gives the dominant contribution \cite{BCGP16} to exchange as $Z\to\infty$,
$E\TF = - A\, Z^{5/3}$.
%
For the beyond-LDA contribution to the exchange energy, \Eqref{eq:beyond_LDA}, we try the GEA \cite{HK64,DG90}, 
\ben
\Delta E\x\GEA = - \mu\GE a\x \int \, \n^{4/3}(\br) \, s^2(\br) \, d^3 r \; ,
\label{GEX}
\een
where $s = |\nabla n|/(2 k\F n)$ is the dimensionless gradient parameter, $k\F = (3 \pi^2)^{1/3} n^{1/3}$ is the local Fermi wavenumber, and
$\mu\GE = 10/81$ \cite{KL88}.
Application of \Eqref{GEX} to the slowly-varying gas\allHL{}{, or to a neutral atom using the density scaling of \cite{PCSB06},} yields a term of order $Z$ when scaled toward the TF limit. 
\allHL{}{However, the \allHL{}{present} analysis 
amounts to scaling
the potential, in the sense of Refs.~\cite{CLEB11,CGB13x}
. 
\allHL{}{While} potential- and density-scaling are interchangeable for the dominant term 
of the large-$Z$ asymptotic expansion (TF theory), 
additional terms 
appear for 
potential scaling, such as the Scott correction to the 
kinetic energy~\cite{LCPB09}.  To show this for exchange, we proceed by directly employing the TF profile in \Eqref{GEX}.}

\begin{figure}
\includegraphics[trim=0cm 0cm 0cm 0cm,clip=true,width=\linewidth]
{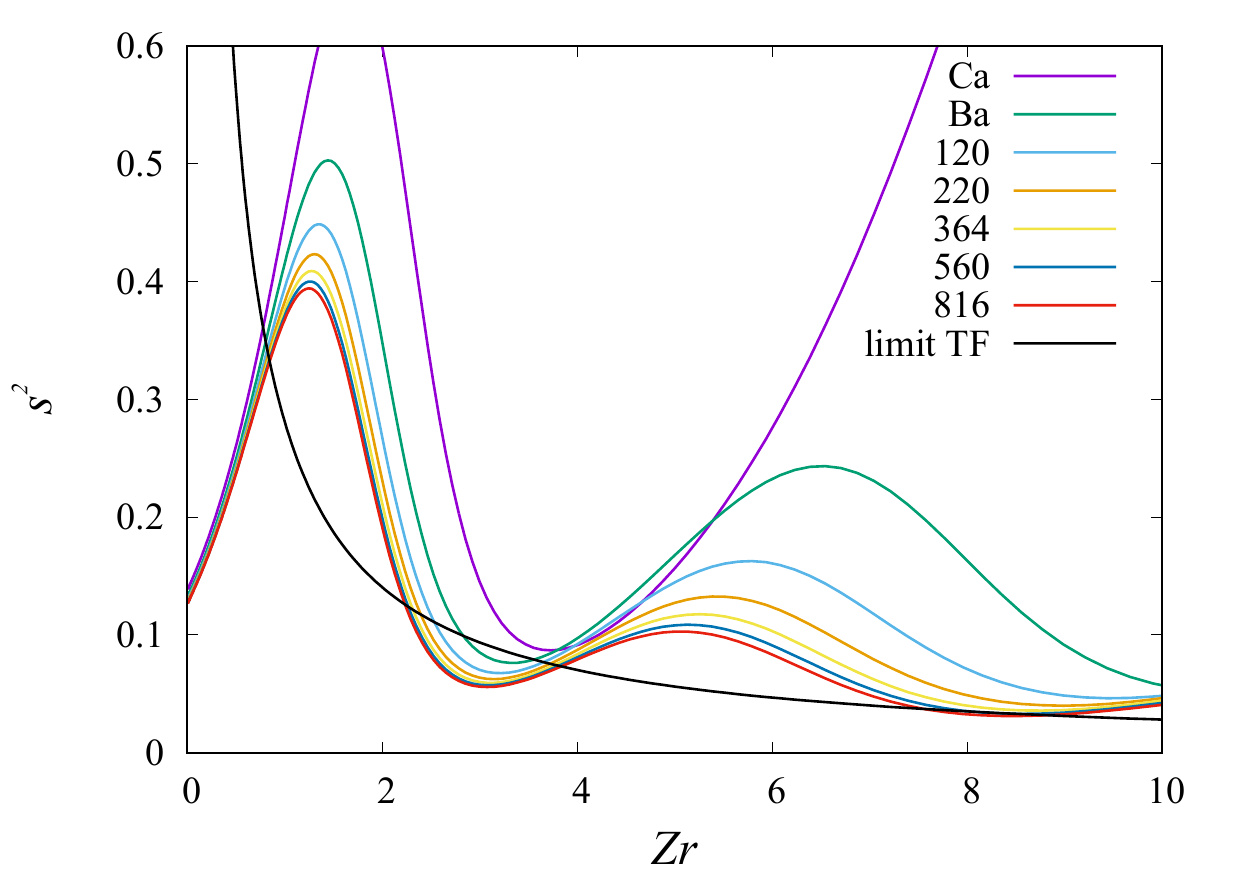}
\caption{Plot of $s^2$ near the nucleus versus distance, scaled as $Zr$, 
for alkaline earth atoms ranging from Ca up to $Z=816$ which has valence shell 16s$^2$. {The black line shows} the TF model. 
\label{fig:s2realatoms}
}
\end{figure}

Gradients are weak in the bulk of large atoms, with $s$ of order $Z^{-1/3}$.
At distances smaller than $O(Z^{-1/3})$ from the nucleus, screening of the nuclear charge is negligible \cite{HL95} 
and the TF density varies as 
$\left({2Z}/{r}\right)^{3/2} /3\pi^2$,
so
\ben
s\TF(r) \simeq \frac{3}{4} \frac{1}{\sqrt{2Zr}}  \; ,
\qquad r \ll Z^{-1/3} \; .
\label{eq:sTF}
\een
This approximation fails in the region where the inner shell (1s) electrons dominate; see Fig.~2, which
shows $s^2$ of alkaline earths up to $Z=816$ (using FHI98PP in all-electron mode \cite{FHI98PP})
and $s^2$ of the TF density, 
\Eqref{eq:sTF}. 
For $r >> 1/Z$, the atomic gradients approach the TF curve, while
near $r \approx 1/Z$, the 
density profile displays the oscillations studied in \cite{HL95} and switches over to that of the well-known nuclear cusp, while $s$ remains finite, achieving its maximum value around $r=1/Z$.  
Keeping only the divergent contribution to \Eqref{GEX} gives:
\begin{equation}
\Delta E\x\GEA \simeq  -\frac{9 \mu\GE}{8 \pi^2} Z 
    \int_{Z^{-1}}^{Z^{-1/3}} \frac{dr}{r} \; ,
\label{DEx}
\end{equation}
which yields a logarithmic term,
\begin{equation}
    \Delta E\x\GEA  = - \frac{3 \mu\GE}{4 \pi^2} Z \ln Z + O(Z) \; . 
\label{GGA_ZlnZ}
\end{equation}
%
We define
\ben
B = -\lim_{Z\to\infty} \Delta E\x/(Z\ln Z) \; ,
\een
and our derivation yields
\ben
B\GEA = \frac{3}{4 \pi^2} \mu\GE
\label{BGEA}
\een
or about $9.38$~mHa.
The presence of such a logarithmic term in the GEA for atoms was noticed in \cite{Paola}, and could be inferred from earlier work (see Appendix A of \cite{CCKB18}).
\allHL{Taking the analog of \Eqref{GEX} for the kinetic energy and repeating our procedure, we again find a divergence at small $r$.  In addition to a term with the naive scaling of $Z^{5/3}$, our cutoff produces a $Z^2$ term, proportional to the Scott correction.  This procedure however cannot generate the needed coefficient, -1/2, which can be calculated~\cite{BCGP16} by considering the Bohr atom, which we do next for exchange.}{}

We have no rationale for the difference between the result of the GEA, \Eqref{BGEA}, and the actual data, \Eqref{eq:fit3}, i.e., the slope in Fig.~1.  
The GEA result is unaffected by integration by parts (unlike 
\cite{PSHPz86}).
Thus, 
the beyond-LDA exchange energy of large-$Z$ atoms has a leading $Z\ln{Z}$ term both numerically and in GEA,
but their coefficients disagree.

\allHL{}{A similar analysis can be applied to the analog of \Eqref{GEX} for the kinetic energy, leading to a stronger divergence at small $r$, due to the presence of an extra power of $n^{1/3}$.  
In addition to the naive-scaling $Z^{5/3}$ term, the small-$r$ cutoff
produces a $Z^2$ term, proportional to the Scott correction mentioned above.  This procedure
does not generate the exact coefficient, -1/2 (see~\cite{LCPB09}).  Instead this is
inferred from the Bohr atom \cite{BCGP16}, to which we turn for the analysis of exchange.}

%
%

The simplicity of the Bohr atom (hydrogenic orbitals) allows calculations to much larger electron number, leading to unambiguous results.  
We fill $N$ hydrogenic orbitals in a potential $-N/r$, so that $N$ plays the role of $Z$ here.
The inner region, $r << N^{-1/3}$, is identical to that of interacting atoms in the large $Z$ limit \cite{KSBW20}.

We analytically evaluated $E\x$, defined by an infinitesimal Coulomb repulsion, up to $N=7590$ (22 shells), 
using Mathematica \cite{SOLR11a}.
Our extremely accurate fit has the form
\bea
&& E\x\Bohr(N) = \\ \nonumber
&& \; -\bar A_o N^{5/3} - (\bar B_o \ln N \! + \bar C_o) N -(\bar D_o \ln N \! + \! \bar E_o) N^{1/3}+...,
\eea
where the subscript denotes a Bohr-atom coefficient and the bar denotes $E\x$.
The leading coefficient is $(2/3)^{1/3} (4/\pi^2)$, from LDA applied to the TF density \cite{KSBW20},
while $\bar B_o$ \allHL{}{$=26.268$ mHa} agrees with $7/(27 \pi^2)$ to 5 digits,  
with $\bar C_o=45.3536$ \allHL{}{mHa}, $\bar D_o=-3.17$ \allHL{}{mHa} and $\bar E_o=0.6$\allHL{, in}{} mHa, determined to the number of digits shown \allHL{}{(see \cite{SM} for details)}. 

For LDA, there are also $O(N^{2/3}\ln N)$ and $O(N^{2/3})$ terms, making
results harder to fit.  However, the simplicity of the expressions \cite{HL95} and availability of arbitrary precision software (using the Julia language with 64-decimal-digit accuracy) enables their brute-force evaluation for up to 100 full shells ($N=676700$) \cite{SM}.  We find $B_o\LDA$ to match $-2/(27 \pi^2)$
to within $\sim 0.1\%$ (note the opposite sign), yielding
\ben
B_o = \bar B_o - B_o\LDA = \frac{1}{3\pi^2} \; .
\label{BBA}
\een

\begin{figure}
\includegraphics[trim=0cm 0cm 0cm 0cm,clip=true,width=\linewidth]
{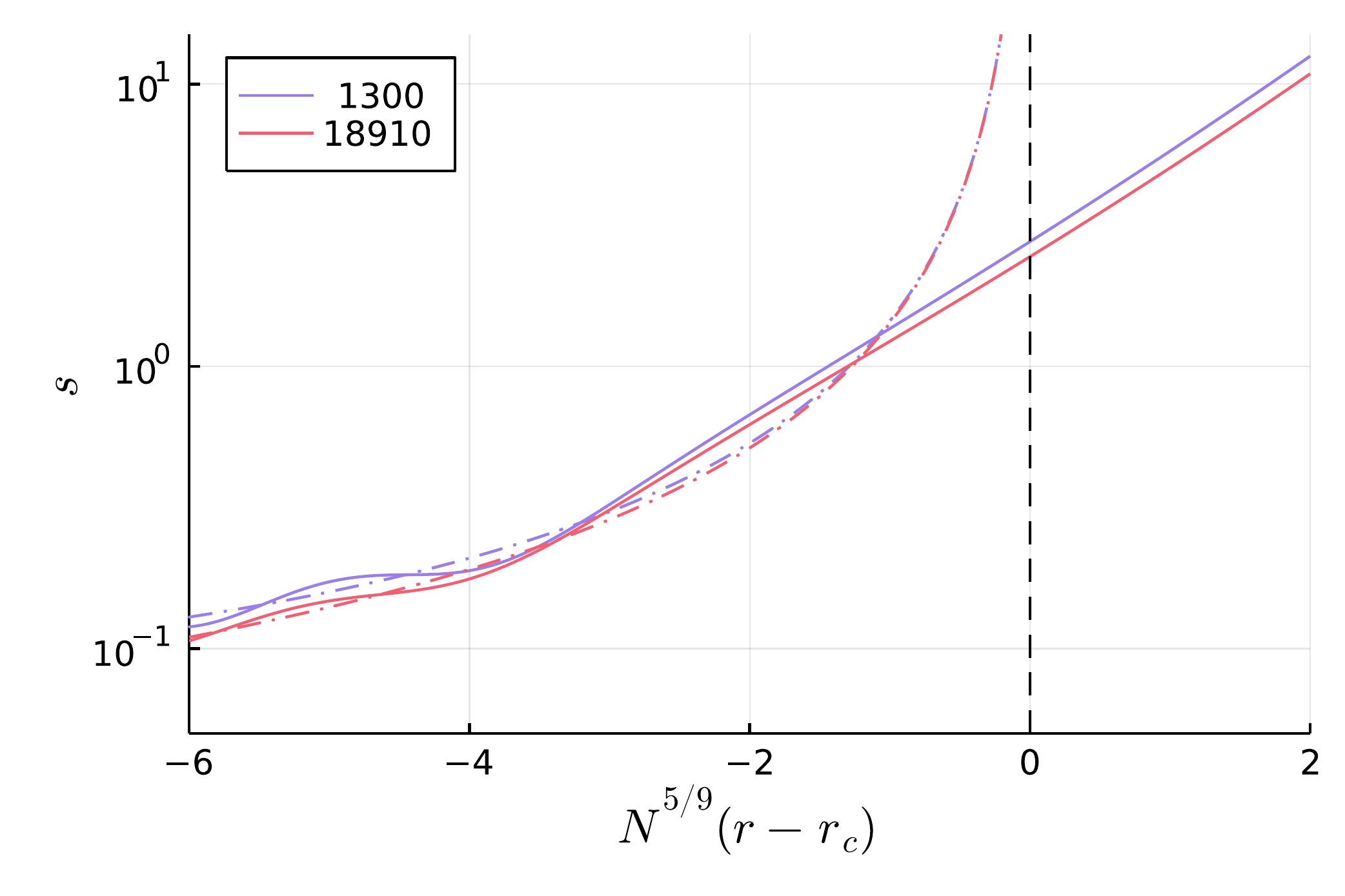}
\caption{Plot of the gradient parameter $s$ near the edge of the Bohr atom, versus distance from the cusp radius $r_c$, scaled by $N^{5/9}$, 
for two representative values of $N$ identified in the legend.  Dot-dashed lines: the results of the corresponding TF models, which diverge at $r_c$ (dashed vertical line).
\label{fig:s_Bohr_atoms}
}
\end{figure}

To evaluate the GEA, note the TF density distribution:
\ben
n\TF_o = \frac{(2N)^{3/2}}{3 \pi^2} \left( r^{-1}-r_c^{-1} \right)^{3/2} \;,
\qquad r \leq r_c \;,
\een
where $r_c=(18/N)^{1/3}$ is the radius
beyond which the density vanishes \cite{KSBW20},
so $s$ diverges not only at the nucleus but also as $r$ approaches $r_c$ \cite{KSBW20}, as 
\ben
s\TF_o \simeq \frac{3}{4} \, \frac{3^{2/3}}{2^{1/6}} \left[ N^{5/9}(r_c-r) \right]^{-3/2} \, ,
\;\;\; 0 < r_c-r \ll N^{-1/3} \;.
\een
The result is 
\begin{equation}
\Delta E_o\GEA \simeq  - \frac{9 \mu\GE}{8 \pi^2} N 
   \left[ \int_{N^{-1}}^{N^{-1/3}} \frac{dr}{r} +
    \int_0^{r_c-N^{-5/9}} \!\!\!\! \frac{dr}{r_c-r} \right] ,
\label{DExB}
\end{equation}
where the first logarithmic divergence is treated as above.  The second is also cut off, taking into account that the kinetic energy is here very small, and the wavelength of the electrons is of order $N^{-5/9}$ \cite{KMd98}, as displayed in Fig.~\ref{fig:s_Bohr_atoms}.
As a result, the contribution of the second divergence is 3 times smaller than that of the first, 
yielding
\ben
B\GEA_o = \frac{\mu\GE}{\pi^2}.
\label{BGEAB}
\een



The two regions of divergence also determine $B_o\LDA$.  
The inner region of the density has been studied in detail in \cite{HL95}.
The leading non-oscillatory correction to the TF density profile
is $n(r) \simeq n\TF(r)[1-1/(64Zr)]$ for $Z^{-1} \ll r \ll Z^{-1/3}$,
producing a contribution of $-1/(18\pi^2)$ via 
\Eqref{EX_LDA}.  
Consistency with the result $B_o\LDA= -2/(27\pi^2)$, \Eqref{BBA}, requires that the outer divergence yields a contribution $1/3$ as large as the first, just as for $B\GEA_o$.


The value excogitated from the highly precise numerical results,
Eq.~(\ref{BBA}), is exactly 27/10 times larger than that of the GEA, \Eqref{BGEAB}. 
It is tempting to conjecture that
\ben
B = \frac{27}{10}\, B\GEA \vphantom{\frac{1}{\binom{1}{2}}}
\een
yields the exact result for {\em all} atoms, including fully interacting
ones, implying that $B=1/(4\pi^2)$ or $25.3$ mHa is the exact result for neutral atoms,
in agreement with our fit, \Eqref{eq:fit3}.   
%
More generally, the conjecture gives the prediction
\ben
B = \frac{1}{12\pi^2} \left(4 - \frac{N}{Z} \right)
\een
for any $N/Z$ ratio, \allHL{}{interpolating between the result for neutral atoms, $N=Z$, and \Eqref{BBA} for $N\ll Z$.}
\allHL{}{
A careful investigation of this relationship will require generating 
data for a large range of $N$ \allHL{}{for each $N/Z$ ratio, as in Fig.~1.} 
\allHL{}{As a preliminary check, we show in~\cite{SM} that} applying this formula with a constant $C$
to a number of positive ions with $N/Z = 1/2$ \allHL{}{continues to give agreement}
with the beyond-LDA data from the OEP, at the $\sim 5\%$ level, for $N>2$.}

Last, we turn to the implications for approximate functional development.
Our derivation applies to most GGA's for the exchange energy, usually written in terms of an enhancement factor $F\x$:
\ben
E\x\GGA = - a\x \int \, \n^{4/3}(\br) F\x\bigl(s(\br)\bigr) \, d^3 r \; .
\label{GGA}
\een
Typically, $F\x \approx 1 + \mu\GGA s^2 +...$ for small $s$, which dominates in the TF limit. 
Thus Eq. (\ref{BGEA}) applies, with $\mu\GE$ replaced by $\mu\GGA$.
This yields $16.7$ mHa for PBE and $20.9$ mHa for B88, 
differing from the value of 25.4 mH of Eq. (\ref{eq:fit3}).  However, both yield accurate $E\x$ for $Z$ between 10 and 100, due to differences in the remaining terms of a large-$Z$ fit.  Thus, functionals that have been fit to large-$Z$ data, such as SCAN, are accurate for all practical calculations. 
In the future both the 
\allHL{
$\ln Z$ terms and higher order corrections should be studied separately}
{$O(Z \ln Z)$ and the $O(Z)$ terms 
should be addressed in developing approximate density functionals.}

Using the hydrogen atom as a `norm' \cite{SRP15}, the conjecture above yields:
\ben
     \Delta E^{\rm normed}\x =  -Z \left\{\frac{\ln Z}{4\pi^2} + \frac{5}{16}-0.2564\right\}
\een
\allHL{}{for neutral atoms,} which is indistinguishable from the straight line of Fig 1, and
contains no empirical parameters.

In conclusion, the present work is a step in the process of improving density functional approximations using asymptotic expansions for non-relativistic atoms: it identifies a logarithmic divergence in the coefficient of the leading $O(Z)$ contribution to the beyond-LDA exchange energy, resulting in a leading $Z\ln Z$ term.

Further steps would involve studying existing approximations, evaluating the coefficients of both their $Z\ln Z$ terms and their $O(Z)$ terms.  Obtaining very-high-$Z$ data for real atoms is crucial, possibly using simplified methods. 
\allHL{}{Analogous data for molecules and solids would also be very helpful, especially to determine any differences
based on the lack of classical turning surfaces in solids \cite{KCBP20}.}
But first and foremost, a derivation of the $Z\ln Z$ term from semiclassical theory, including the correct value of its coefficient, would provide a fundamental, detailed understanding of the exchange energy, and would be instrumental in guiding future developments in density functional theory.

This research was supported by the NSF \allHL{}{(CHE-2154371)}.  We thank Paola
Gori-Giorgi for {communicating a preprint of \cite{Paola} and
for the ensuing valuable discussions, and John Snyder
for unpublished notes.}

\bibliography{Master,burkegroup,additions}

\label{page:end}

\end{document}


\sf 
\coloredtitle{
Supplementary Material for 
``Leading correction to the local density approximation for exchange in large-$Z$ atoms''
}
\author{\color{CITECOL} Nathan Argaman}
\affiliation{Department of Physics, Nuclear Research Center---Negev, P.O. Box 9001, Be'er Sheva 84190, Israel; argaman@mailaps.org}
\author{\color{CITECOL} Jeremy Redd}
\affiliation{Department of Physics, Utah Valley University, Orem, UT 84058, USA}
\author{\color{CITECOL} Antonio C. Cancio}
\affiliation{Department of Physics and Astronomy, Ball State University,
Muncie, IN 47306, USA}
\author{\color{CITECOL} Kieron Burke}
\affiliation{Departments of Physics and Astronomy and of Chemistry, 
University of California, Irvine, CA 92697,  USA}

\date{4 October, 2022}

\begin{abstract}
{Tables of exchange-energy data for neutral atoms, positive ions and Bohr atoms are given, as well as details of the fits to asymptotic expansions in the number of electrons.}
\end{abstract}

\maketitle


\section{Energies for neutral atoms}

In Table~\ref{atomdata} we tabulate the per-electron exchange energies
$\epsilon\x=E\x/Z$ (in Ha/electron) for neutral atoms with $1\leq Z\leq 120$.
We include data for the spin-dependent local density approximation (LDA) and the optimized effective potential (OEP). 
\begin{table*}
\begin{ruledtabular}
\begin{tabular}{l r r | l r r | c c c }
Z   &  LDA  & OEP & Z & LDA & OEP & Z & LDA & OEP \\
\hline
1 & -0.25643 & -0.31250 & 41 & -2.66072 & -2.81060 & 81 & -4.18152 & -4.34816\\
2 & -0.43087 & -0.51288 & 42 & -2.70355 & -2.85462 & 82 & -4.21322 & -4.38052\\
3 & -0.50477 & -0.59358 & 43 & -2.74045 & -2.89248 & 83 & -4.24477 & -4.41280\\
4 & -0.57258 & -0.66644 & 44 & -2.78341 & -2.93574 & 84 & -4.27513 & -4.44358\\
5 & -0.64936 & -0.74854 & 45 & -2.82457 & -2.97753 & 85 & -4.30552 & -4.47437\\
6 & -0.73837 & -0.84105 & 46 & -2.86983 & -3.02421 & 86 & -4.33591 & -4.50526\\
7 & -0.83672 & -0.94349 & 47 & -2.90896 & -3.06390 & 87 & -4.36394 & -4.53342\\
8 & -0.91253 & -1.02256 & 48 & -2.94660 & -3.10166 & 88 & -4.39146 & -4.56096\\
9 & -0.99989 & -1.11145 & 49 & -2.98276 & -3.13854 & 89 & -4.41983 & -4.58942\\
10 & -1.09667 & -1.21050 & 50 & -3.01882 & -3.17524 & 90 & -4.44834 & -4.61803\\
11 & -1.15723 & -1.27391 & 51 & -3.05481 & -3.21198 & 91 & -4.48309 & -4.65237\\
12 & -1.21362 & -1.33236 & 52 & -3.08888 & -3.24639 & 92 & -4.51534 & -4.68464\\
13 & -1.26815 & -1.38945 & 53 & -3.12318 & -3.28100 & 93 & -4.54795 & -4.71740\\
14 & -1.32461 & -1.44808 & 54 & -3.15765 & -3.31599 & 94 & -4.58464 & -4.75454\\
15 & -1.38288 & -1.50894 & 55 & -3.18824 & -3.34653 & 95 & -4.61813 & -4.78864\\
16 & -1.43435 & -1.56220 & 56 & -3.21811 & -3.37619 & 96 & -4.64777 & -4.81868\\
17 & -1.48856 & -1.61782 & 57 & -3.24965 & -3.40767 & 97 & -4.68219 & -4.85217\\
18 & -1.54512 & -1.67637 & 58 & -3.29489 & -3.45176 & 98 & -4.71470 & -4.88459\\
19 & -1.58748 & -1.71931 & 59 & -3.33473 & -3.49126 & 99 & -4.74752 & -4.91745\\
20 & -1.62795 & -1.75995 & 60 & -3.37534 & -3.53175 & 100 & -4.78064 & -4.95077\\
21 & -1.67794 & -1.81033 & 61 & -3.41663 & -3.57320 & 101 & -4.81404 & -4.98456\\
22 & -1.73095 & -1.86381 & 62 & -3.45858 & -3.61558 & 102 & -4.84774 & -5.01882\\
23 & -1.79706 & -1.92995 & 63 & -3.50114 & -3.65889 & 103 & -4.87607 & -5.04771\\
24 & -1.85542 & -1.98981 & 64 & -3.53680 & -3.69514 & 104 & -4.90703 & -5.07926\\
25 & -1.90284 & -2.03933 & 65 & -3.58066 & -3.73645 & 105 & -4.93670 & -5.10953\\
26 & -1.95511 & -2.09125 & 66 & -3.62133 & -3.77647 & 106 & -4.96638 & -5.13987\\
27 & -2.01567 & -2.15052 & 67 & -3.66260 & -3.81731 & 107 & -4.99609 & -5.17034\\
28 & -2.07288 & -2.20820 & 68 & nan & -3.87481 & 108 & -5.02461 & -5.19913\\
29 & -2.13180 & -2.26810 & 69 & -3.74685 & -3.90154 & 109 & -5.05324 & -5.22806\\
30 & -2.18308 & -2.32063 & 70 & -3.78980 & -3.94495 & 110 & -5.08197 & -5.25715\\
31 & -2.23125 & -2.37083 & 71 & -3.82682 & -3.98320 & 111 & -5.11078 & -5.28643\\
32 & -2.27892 & -2.42026 & 72 & -3.86369 & -4.02129 & 112 & -5.13969 & -5.31593\\
33 & -2.32634 & -2.46958 & 73 & -3.90045 & -4.05929 & 113 & -5.16617 & -5.34266\\
34 & -2.37021 & -2.51475 & 74 & -3.93716 & -4.09730 & 114 & -5.19258 & -5.36928\\
35 & -2.41456 & -2.56025 & 75 & -3.97388 & -4.13544 & 115 & -5.21892 & -5.39588\\
36 & -2.45932 & -2.60647 & 76 & -4.00870 & -4.17097 & 116 & -5.24447 & -5.42149\\
37 & -2.49740 & -2.64514 & 77 & -4.04369 & -4.20670 & 117 & -5.27005 & -5.44711\\
38 & -2.53421 & -2.68227 & 78 & -4.08121 & -4.24554 & 118 & -5.29563 & -5.47281\\
39 & -2.57363 & -2.72217 & 79 & -4.11684 & -4.28235 & 119 & -5.31963 & -5.49667\\
40 & -2.61419 & -2.76329 & 80 & -4.14968 & -4.31556 & 120 & -5.34330 & -5.52012\\
\end{tabular}
\caption{Exchange energy per electron for neutral atoms for $Z=1$ through $Z=120$, using
the PW92 local density approximation (LDA) and the optimized effective 
potential (OEP).  \label{atomdata}
}
\end{ruledtabular}
\end{table*}

\section{Energies for Positive Ions}

Table~\ref{table:ions} gives the data for selected positive ions, focusing on a series 
with the ratio of electron number to nuclear charge, $N/Z$, set to one-half.
The beyond-LDA exchange energy per particle,
$\Delta E\x/Z = (E\x - E\x\LDA)/Z$, is included, and compared with the model suggested 
in the text: $- B(N/Z) \ln{N} - C$, 
with $B(N/Z)$ given by Eq.~(18), and $-C$ equal to the beyond
LDA value of the Hydrogen atom.  
The ratio $N/Z=1/2$ probes a situation roughly halfway between
the neutral atom and the Bohr atom, the two limits 
of Eq.~(18) studied in detail.
The $\ln{Z}$ term has been altered to $\ln{N}$ in order to
produce reasonable results in the Bohr-atom limit and for single-electron systems (also included in the table). 
Replacing $\ln N$ by $\ln Z$ is equivalent to shifting $C$ by
$B\ln{N/Z}$, which is a constant for fixed $N/Z$.  
{With the use of $\ln N$, the value of $C$ used is $3\%$ off from that for the Bohr atom (see the end of Sec.~4 below).}


%

\begin{table*}[h]
\begin{ruledtabular}
\begin{tabular}{ r | r | c | c | c | c | r }
$N$ & $Z$ & OEP & LDA &  $\Delta E\x/Z$ & model & \% difference \\
\hline
1 & 2 & -0.3125 & -0.2616 & 	-0.0509 & -0.0561 & -10.2 \\
2 & 4 & -0.5693 & -0.4829 &	-0.0864 & -0.0766 & 11.4 \\
4 & 8 & -0.7443 & -0.6461 & 	-0.0982 & -0.0971 & 1.1 \\
10 & 20 & -1.4872 & -1.3591 & 	-0.1280 & -0.1241 & 3.0 \\
12 & 24 & -1.6026 & -1.4736 & 	-0.1290 & -0.1295 & -0.4 \\
18 & 36 & -2.0145 & -1.8771 & 	-0.1374 & -0.1415 & -3.0 \\
1 & 1 & -0.3125 & -0.2564 & 	-0.0561 & -0.0561 & -0.0 \\
1 & 2 & -0.3125 & -0.2616 & 	-0.0509 & -0.0561 & -10.2 \\
1 & 4 & -0.3125 & -0.2646 & 	-0.0479 & -0.0561 & -17.2 \\
1 & 10 & -0.3125 & -0.2666 & 	-0.0459 & -0.0561 & -22.3 \\
1 & 12 & -0.3125 & -0.2669 & 	-0.0456 & -0.0561 & -22.9 \\
1 & 18 & -0.3125 & -0.2672 & 	-0.0453 & -0.0561 & -24.0 \\
%
\end{tabular}
\end{ruledtabular}
\caption{Exchange energies {divided by $Z$} for various positive ions.   Shown are exact exchange using the OEP method, the LDA,  the beyond-LDA contribution as compared to an asymptotic model, and the percent error of the model.}
\label{table:ions}
\end{table*}

\section{Statistical Fits}
In Table~\ref{table:fits} we show systematic
data fits for $\Delta E\x /Z=$  ${(E\x - E\x^{LDA})/Z}$ for the neutral atoms of Table~\ref{atomdata}.  The first column
indicates the data set used, as explained below, then 
coefficients with asymptotic standard errors from nonlinear regression,
using the Levenberg-Marquardt method.  Finally, the reduced $\chi^2$ which
is the $\chi^2$ measure divided by the net number of degrees of freedom
in the fit.  
{In calculating the reduced $\chi^2$, a standard error of 1~mHa is assumed
for individual energy data points.}

There are four data sets used here.  
``all" uses all atoms from $Z=1$ to 120 indiscriminately.  
The large data set, ``l", consists of 16 data points, corresponding to atoms with closed s, p, d, and f shells, excluding the first occurrence of each
series, He (1s$^2$), Ne (2p$^6$), Zn (3d$^{10}$) and Yb (4f$^{14}$). 
The atoms in the set thus consist of the filled 2s through 8s ($Z=120$) alkali earths,
3p through 7p ($Z=118$) noble gases, 4d through 6d {group 12 transition}
metals and the filled 5f actinide.
The net number
of degrees of freedom varies from 12 to 15, depending on the number of fit parameters.
The medium data set, labelled ``m", drops the next smallest closed shell of each series, 
the closed 2s, 3p, 4d and 5f
shell atoms, and thus has 12 atoms.
The ``s" or small data set drops the next smallest shell (3s, 4p, 5d), for 9 atoms.

The basic model used for all fits is [Eq.~(2) of  main text]:
$$
\Delta E\x (Z) /Z \approx -(A^\prime Z^{1/3} + B\log{Z} + C + DZ^{-1/3})
$$
which is fit versus $x\!=\!Z^{1/3}$, so the actual fit equation used is:
$$
y = -A^\prime x - 3B \log{x} - C - D/x.
$$

Assuming $C\neq 0$, there are eight possible models formed by
setting $A$, $B$ or $D$ to be either zero (in which case the data entry
is left blank) or nonzero.  All eight are shown here for completeness,
but in the main text, the fifth, which is noncompetitive is omitted.



\begin{table}[htp]
\begin{ruledtabular}
\begin{tabular}{l|c|c|c|c|r}
data \\ set & $A^\prime$  & $B$ & $C$ & $D$ & $\chi^2_{red}$ \\
\hline 

all & & & 0.1516(21)   & & 530\\
l & & & 0.153(6)     & & 560\\
m & & & 0.158(5)     & & 359\\
s & & & 0.163(5)     & & 240\\
\hline

all & & & 0.2048(16)    & -0.179(5)    & 41.7\\
l & & & 0.2138(34)   & -0.205(11)     & 22.1\\
m & & & 0.2269(25)   & -0.256(9)      & 4.8\\
s & & & 0.2328(24)   & -0.279(9)     & 2.1\\
\hline

all & & 0.02432(24)   & 0.0589(9)    & & 5.8\\
l & & 0.02464(26)  & 0.0590(10)   & & 0.91\\
m & & 0.02538(26)   & 0.0560(11)   & & 0.40\\
s & & 0.02535(32)   & 0.0562(13)   & & 0.29\\
\hline

all & & 0.0260(9)  & 0.049(6)   & 0.013(7)    & 5.7 \\
l & & 0.0256(14)   & 0.053(9)      & 0.008(12)     & 0.95\\
m & & 0.0238(23)   & 0.0667(16)   & -0.016(23)     & 0.42\\
s & & 0.0225(35)   & 0.076(24)    & -0.03(4)     & 0.30\\
\hline

all & 0.0238(5)  & & 0.0630(17)   & & 21. \\
l & 0.0230(10)   & & 0.067(4)     & & 16. \\
m & 0.021(8)   & & 0.0747(33)     & & 5.8 \\
s & 0.0199(8)  & & 0.0807(34)  & & 3.0\\
\hline

all & 0.0149(5)  & & 0.1189(30)   & -0.0759(39)   & 5.1 \\
l & 0.0128(9)    & & 0.134(5)     & -0.098(7)       & 1.3 \\
m & 0.0103(10)  & & 0.154(7)     & -0.136(12)     & 0.43 \\
s & 0.0090(14)  & & 0.165(11)     & -0.155(20)      & 0.33 \\
\hline

all &  0.0032(11)   & 0.0212(12)   & 0.0590(9)  &    & 5.5 \\
l &  0.0007(15)    &  0.0239(16)    & 0.0592(11)    &   & 0.96\\
m &  -0.0013(20)     & 0.0285(35)     & 0.0549(20)    &     & 0.42\\
s &  -0.0025(28)     &  0.0269(24)   & 0.0533(36)   &     & 0.30\\
\hline

all &  0.014(4)   & 0.008(7)    & 0.117(19)      & -0.073(24)    & 5.1 \\
l &  -0.007(8)    & 0.039(16)     & 0.01(5)       & 0.06(7)       & 0.98\\
m &  0.003(17)     & 0.02(4)      & 0.09(15)      & -0.05(20)       & 0.47\\
s &  -0.025(35)    & 0.08(8)      & -0.16(34)     & 0.3(5)     & 0.33\\

\end{tabular}
\end{ruledtabular}
\caption{Coefficients and statistics for data fits to neutral atoms.
Coefficients match those of Eq.~(2) and Table~I of the main text.\label{table:fits}}
\end{table}

\clearpage
\section{The Bohr atom}

The exchange energies for the Bohr atom were fit by defining a residual
$$ R_o(N) = [E\x\Bohr(N) + \bar A_o N^{5/3}]/N \: , $$
in lieu of $E\x\Bohr(N)$ itself (recall that $\bar A_o = (2/3)^{1/3} (4/\pi^2)$).  The values of this residual are provided in Table~\ref{BohrEx} for up to $n=22$ full shells, and are seen to vary nearly linearly in $\ln N$, with the deviations from linearity decreasing for large $N$.  In order to obtain many-digit accuracy for the coefficients, a second residual,
$$ S_o(N) = [R_o(N) + \bar B_o \ln N + \bar C_o] N^{2/3} \: , $$
was defined, and it too varies nearly linearly in $\ln N$.  
The most accurate fit was obtained by inspecting visually plots of 
$S_o(N) + \bar D_o \ln N$ vs.\ $\ln N$, magnifying the deviation of the second residual from linearity in $\ln N$, and adjusting the values of the coefficients so that the deviations from linearity at large $N$ are minimal [once it was guessed that $\bar B_o = 7/(27 \pi^2)$, the analytic value was used for subsequent refinement, so that no more than two coefficients needed to be simultaneously adjusted].  Obtaining smooth plots requires retaining more than 6 significant digits in $R_o(N)$, due to the multiplication by $N^{2/3}$ and the magnification (even more significant digits are required in $E\x\Bohr(N)$, of course).

\begin{table}[b]

\begin{ruledtabular}
\begin{tabular}{c c c c c}
& $n$ & $N$ & $R_o(N)$ &\\
\hline
& 1 & 2 & -0.06298252 \\  
& 2 & 10 & -0.10453258 \\ 
& 3 & 28 & -0.13185039 \\
& 4 & 60 & -0.15211566 \\
& 5 & 110 & -0.16821274 \\
& 6 & 182 & -0.18156344 \\
& 7 & 280 & -0.19296926 \\
& 8 & 408 & -0.20292569 \\
& 9 & 570 & -0.21176023 \\ 
& 10 & 770 & -0.21970057 \\
& 11 & 1012 & -0.22691143 \\
& 12 & 1300 & -0.23351580 \\
& 13 & 1638 & -0.23960797 \\
& 14 & 2030 & -0.24526180 \\
& 15 & 2480 & -0.25053622 \\
& 16 & 2992 & -0.25547902 \\
& 17 & 3570 & -0.26012950 \\
& 18 & 4218 & -0.26452035 \\
& 19 & 4940 & -0.26867906 \\
& 20 & 5740 & -0.27262896 \\
& 21 & 6622 & -0.27639005 \\
& 22 & 7590 & -0.27997958 \\
\end{tabular}
\caption{The residual of the exchange energy per electron for Bohr atoms with $n$ complete shells.  $N$ is the number of electrons. \label{BohrEx}
}
\end{ruledtabular}
\end{table}

For the LDA applied to the Bohr atom, a residual $R_o\LDA(N)$ was similarly defined, with values given in Table~\ref{Bohr_LDA}.  In this case the deviations from linearity are greater, due to the presence of additional terms in the expansion --- here, the second residual would be defined with a power of $N^{1/3}$ rather than $N^{2/3}$.  The similarity of the $\propto N$ and the $\propto N^{2/3}\ln N$ behaviors over a large range of $N$ makes fitting by visual inspection difficult (the accuracy achieved for $\bar B_o\LDA$ based on data up to $n=28$ shells was circa 1\%, leaving some room for questions regarding the use of the analytic value).  Luckily, extension of the data set to very large values of $N$ is accessible, up to $n=100$ shells here, and an automated fit provides sufficient accuracy (which is gauged by comparison to fits with more limited ranges of data).

{The fit gives $\bar B_o\LDA = -7.505$ mHa and  $\bar C_o\LDA = -9.2$ mHa (further coefficients were not carefully extracted).  Extracting the beyond-LDA coefficient as in Eq.~(12) gives $C_o = 54.6$ mHa for the Bohr atom, which differs by only a few percent from the neutral-atom value in Eq.~(3).}


\begin{table*}[htp]
\begin{ruledtabular}
\begin{tabular}{c c c | c c c | c c c | c c c }
$n$ & $N$ & $R_o\LDA(N)$ & $n$ & $N$ & $R_o\LDA(N)$ & $n$ & $N$ & $R_o\LDA(N)$ & $n$ & $N$ & $R_o\LDA(N)$\\
\hline
1 & 2 & 0.02594248 & 26 & 12402 & 0.08262241 & 51 & 91052 & 0.09663426 & 76 & 298452 & 0.10511544 \\ 
2 & 10 & 0.03499465 & 27 & 13860 & 0.08339534 & 52 & 96460 & 0.09704426 & 77 & 310310 & 0.10539532 \\
3 & 28 & 0.04156255 & 28 & 15428 & 0.08414161 & 53 & 102078 & 0.09744674 & 78 & 322478 & 0.10567170 \\ 
4 & 60 & 0.04657061 & 29 & 17110 & 0.08486301 & 54 & 107910 & 0.09784198 & 79 & 334960 & 0.10594467 \\ 
5 & 110 & 0.05059082 & 30 & 18910 & 0.08556118 & 55 & 113960 & 0.09823025 & 80 & 347760 & 0.10621431 \\
6 & 182 & 0.05394465 & 31 & 20832 & 0.08623759 & 56 & 120232 & 0.09861179 & 81 & 360882 & 0.10648071 \\
7 & 280 & 0.05682234 & 32 & 22880 & 0.08689356 & 57 & 126730 & 0.09898682 & 82 & 374330 & 0.10674393 \\
8 & 408 & 0.05934382 & 33 & 25058 & 0.08753032 & 58 & 133458 & 0.09935556 & 83 & 388108 & 0.10700407 \\
9 & 570 & 0.06158899 & 34 & 27370 & 0.08814897 & 59 & 140420 & 0.09971824 & 84 & 402220 & 0.10726118 \\
10 & 770 & 0.06361357 & 35 & 29820 & 0.08875052 & 60 & 147620 & 0.10007504 & 85 & 416670 & 0.10751533 \\
11 & 1012 & 0.06545792 & 36 & 32412 & 0.08933592 & 61 & 155062 & 0.10042615 & 86 & 431462 & 0.10776661 \\
12 & 1300 & 0.06715218 & 37 & 35150 & 0.08990601 & 62 & 162750 & 0.10077175 & 87 & 446600 & 0.10801506 \\
13 & 1638 & 0.06871947 & 38 & 38038 & 0.09046158 & 63 & 170688 & 0.10111202 & 88 & 462088 & 0.10826076 \\
14 & 2030 & 0.07017791 & 39 & 41080 & 0.09100336 & 64 & 178880 & 0.10144712 & 89 & 477930 & 0.10850376 \\
15 & 2480 & 0.07154194 & 40 & 44280 & 0.09153204 & 65 & 187330 & 0.10177720 & 90 & 494130 & 0.10874413 \\
16 & 2992 & 0.07282331 & 41 & 47642 & 0.09204822 & 66 & 196042 & 0.10210242 & 91 & 510692 & 0.10898191 \\
17 & 3570 & 0.07403166 & 42 & 51170 & 0.09255251 & 67 & 205020 & 0.10242291 & 92 & 527620 & 0.10921717 \\
18 & 4218 & 0.07517502 & 43 & 54868 & 0.09304543 & 68 & 214268 & 0.10273881 & 93 & 544918 & 0.10944996 \\
19 & 4940 & 0.07626018 & 44 & 58740 & 0.09352751 & 69 & 223790 & 0.10305025 & 94 & 562590 & 0.10968033 \\
20 & 5740 & 0.07729287 & 45 & 62790 & 0.09399920 & 70 & 233590 & 0.10335737 & 95 & 580640 & 0.10990833 \\
21 & 6622 & 0.07827803 & 46 & 67022 & 0.09446095 & 71 & 243672 & 0.10366027 & 96 & 599072 & 0.11013401 \\
22 & 7590 & 0.07921992 & 47 & 71440 & 0.09491317 & 72 & 254040 & 0.10395907 & 97 & 617890 & 0.11035741 \\
23 & 8648 & 0.08012225 & 48 & 76048 & 0.09535626 & 73 & 264698 & 0.10425388 & 98 & 637098 & 0.11057859 \\
24 & 9800 & 0.08098825 & 49 & 80850 & 0.09579058 & 74 & 275650 & 0.10454481 & 99 & 656700 & 0.11079758 \\
25 & 11050 & 0.08182080 & 50 & 85850 & 0.09621647 & 75 & 286900 & 0.10483197 & 100 & 676700 & 0.11101443 \\
\end{tabular}
\caption{The residual for the Bohr atom within the LDA.  \label{Bohr_LDA}
}
\end{ruledtabular}
\end{table*}
